\documentclass[%
twocolumn,
amsmath,
amssymb,
showpacs,
 aps,
 prl,
 lengthcheck,%
]{revtex4-1}
\usepackage{graphicx}
\usepackage{dcolumn}
\usepackage{bm}
\usepackage{hyperref}
\usepackage{makeidx}

\usepackage[english]{babel} 
\usepackage[utf8]{inputenc}
\usepackage{color}

\makeindex

\newcommand{\cS}{{\cal S}}

\newcommand{\cI}{{\cal I}}
\newcommand{\cT}{{\cal T}}
\newcommand{\cG}{{\cal G}}
\newcommand{\cC}{{\cal C}}

\newcommand{\bT}{\bar{T}}
\newcommand{\DT}{\Delta T}

\begin{document}
 
\title{Negative delta-$T$ noise in the Fractional Quantum Hall effect}

\author{J. Rech, T. Jonckheere, B. Grémaud, and T. Martin}
\affiliation{Aix Marseille Univ, Université de Toulon, CNRS, CPT, Marseille, France}

\date{\today}

\begin{abstract}
We study the current correlations of fractional quantum Hall edges at the output of a quantum point contact (QPC) subjected to a temperature gradient. This out-of-equilibrium situation gives rise to a form of temperature-activated shot noise, dubbed delta-$T$ noise. We show that the tunneling of Laughlin quasiparticles leads to  a \emph{negative} delta-$T$ noise, in stark contrast with electron tunneling. Moreover, varying the transmission of the QPC or applying a voltage bias across the Hall bar may flip the sign of this noise contribution, yielding signatures which can be accessed experimentally.
\end{abstract}

\maketitle


\emph{Introduction.---}
Noise is a fundamentally inescapable ingredient of any electronic device. While at first it may be regarded as a nuisance, it has now been broadly accepted as a key tool to improve our understanding of nanoscale conductors. Electronic noise is typically broken down into two contributions associated with different underlying physical phenomena. Thermal (or Johnson-Nyquist) noise is an equilibrium property, arising at finite temperature from the thermal motion of electrons~\cite{johnson27, nyquist28}. Shot noise manifests itself in a non-equilibrium situation, when current flows through a conductor, as a consequence of electrons being transmitted or reflected predominantly on a given side of the device~\cite{schottky18}. 

Shot noise has taken a massive role in quantum mesoscopic physics, where it is commonly used to extract a great deal of information on the mechanisms of charge transfer (for a review see e.g. \cite{blanter00, martin05}). While this non-equilibrium situation is typically achieved by imposing a bias voltage on the device, an intriguing alternative was recently uncovered. Indeed, one can in principle work at zero voltage bias and instead connect the sample to two reservoirs at different temperatures. This was realized experimentally using atomic-scale metallic junctions~\cite{lumbroso18}, where the authors showed that, while no net current was flowing through the device, as expected, a finite non-equilibrium noise signal was measured, which they dubbed ``delta-$T$ noise''. This previously undocumented source of noise, distinct from thermoelectric effects, actually corresponds to some form of temperature-activated shot noise: it is purely thermal in origin, but only generated in a non-equilibrium situation. Its properties also present the same hybrid character, as this positive contribution to the total noise scales like the square of the temperature difference, while exhibiting the same dependence on the conductance as shot noise. Temperature-activated shot noise was exploited earlier to realize a local noise measurement in a current biased metallic conductor~\cite{tikhonov16}.

As it turns out, delta-$T$ noise is accurately described by the standard quantum theory of charge transport~\cite{lumbroso18}. Remarkably, the scattering theory~\cite{landauer57} does predict the appearance of this overlooked source of noise when a temperature difference alone is applied across a metallic junction. However,  while the Fermi statistics of the charge carriers is accounted for within this formalism, interaction effects between electrons are discarded. We show here that delta-$T$ noise can be a unique tool to explore the properties of the charge carriers in a system where interactions and statistics play a significant role.

\begin{figure}[tb]
\includegraphics[width=8.6cm]{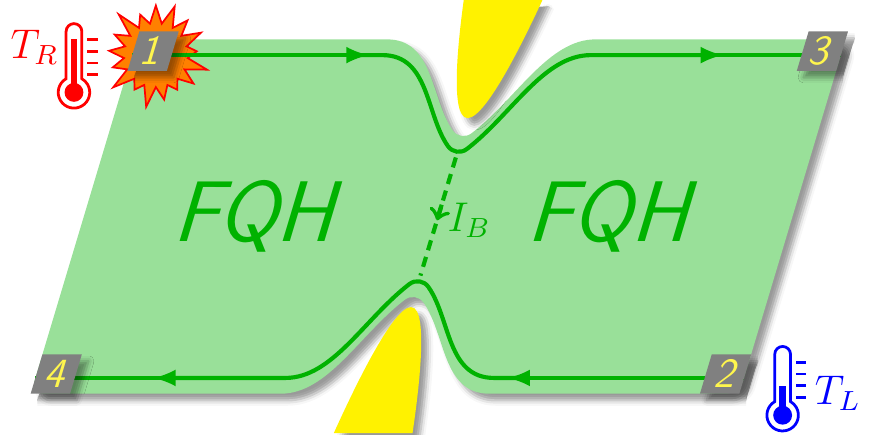}
\caption{Main setup: a quantum Hall bar equipped with a QPC connecting the chiral edge states of the FQH. The left-moving incoming edge is grounded at contact $2$, with a temperature $T_L$ while the right-moving one is heated by contact $1$, reaching a temperature $T_R$. Relevant output currents are measured from contacts $3$ and $4$.
	}
\label{fig:setup}
\end{figure}

We propose to investigate the fate of delta-$T$ noise in a prototypical strongly correlated state, namely the edge states of the fractional quantum Hall effect (FQHE)~\cite{laughlin99, stormer99, tsui99}. In this system, excitations propagate along chiral edge modes, while the bulk is gapped. Transport through fractional quantum Hall edges has been extensively studied over the past few decades, both theoretically and experimentally. This interest is in part related to the bulk-boundary correspondence, as edge transport reflects the topological nature of the bulk, revealing nontrivial properties unreachable through direct measurements. In particular, noise measurements in quantum point contact geometries (QPC) have been used experimentally to identify the fractional charge of the quasiparticle excitations in the bulk~\cite{saminadayar97, depicciotto97, dolev08}, and several proposals of more elaborate interferometric devices have been put forward as a way to probe their fractional statistics~\cite{chamon97,safi01,vishveshwara03,kane03,kim05,camino05,law06,campagnano12,campagnano13} with promising recent experimental results~\cite{bartolomei20, nakamura20}. These considerations make the edge states of the FQHE a perfect testbed to examine delta-$T$ noise.
Beyond the inherent interest in studying delta-$T$ noise in such systems, it may help better understanding charge and heat transport in situations where strong electronic correlations are operating~\cite{sivre19}. The present work is an essential first step before more involved setups are considered, including e.g. edge states involving neutral modes~\cite{bid10,shtanko14} or more exotic fractions, where intriguing transport properties associated with thermal effects have been uncovered recently~\cite{banerjee17, banerjee18}.

Here we show that, for Laughlin fractions, the delta-$T$ noise signal is \emph{negative} in the weak backscattering regime, a result which arises from the interplay of strong correlations and fractional statistics, and is directly accessible in modern experiments. 

\emph{Delta-$T$ noise.---}
We start by recalling results concerning delta-$T$ noise in non-interacting fermionic systems~\cite{lumbroso18}.  Consider a system of two fermionic reservoirs (with Fermi distribution $f_1$ and $f_2$ respectively) separated by a barrier of transmission $\cT$. The fluctuations $\delta \hat{I} = \hat{I} -  \langle \hat{I} \rangle$ of the current operator away from its average value are monitored using the zero-frequency current noise, defined as $S = 2\int dt \langle \delta \hat{I} (t) \delta \hat{I} (0) \rangle$. Within scattering theory, this quantity assumes a very general form~\cite{blanter00}
\begin{align}
S = \frac{e^2}{\pi} \int dE &\left\{ \cT \left[ f_1 \left( 1 - f_1 \right) + f_2 \left( 1 - f_2 \right) \right] \right. \nonumber \\
& \left. + \cT \left( 1- \cT \right) \left( f_1 - f_2 \right)^2  \right\}
\end{align}
where we considered for simplicity a single conduction channel, and approximated the transmission coefficient by assuming it energy-independent (setting $\hbar=1$ from this point onward). The first term corresponds to thermal-like noise, while the last one is a non-equilibrium contribution. Let us now focus on the situation where no voltage bias is applied but the reservoirs are at different temperatures, yielding a temperature difference $\DT$, with an average temperature $\bT$. Expanding for a small temperature difference leads to the following approximate expression for the current noise~\cite{lumbroso18}
\begin{align}
S \approx 2 \frac{e^2}{\pi}  &\left[ \cT \bT + \cT \left( 1- \cT \right) \frac{\pi^2 - 6}{9} \left( \frac{\DT}{2 \bT} \right)^2 \bT  \right]
\end{align}
where, along with the equilibrium thermal noise, another non-equilibrium component arises as a result of the temperature difference across the junction: delta-$T$ noise~\footnote{See Supplemental Material for further details of the derivation.}.

While these results apply to a very broad range of non-interacting devices, scattering theory is expected to fail in the fractional case calling for another formalism. It should, however, provide an accurate description of the properties of the system in the Fermi liquid regime, i.e. at filling factor $\nu =1$, or for a two-dimensional electron gas in the absence of a strong magnetic field (like the setup considered in \cite{dubois13} in the context of electron quantum optics). Our present derivation bridges the gap, in providing a formalism that not only recovers the above Fermi-liquid result in the proper limit, but also extends it to the case of quasiparticle tunneling.

\emph{Model.---}
The system considered here is a Hall bar (see Fig.~\ref{fig:setup}), in the fractional quantum Hall regime, with a filling factor in the Laughlin sequence, i.e. $\nu = 1/(2 n +1)$ ($n\in\mathbb N$). The edge states of such a sample are described in terms of a hydrodynamical model~\cite{wen95} by a chiral Luttinger liquid Hamiltonian of the form
\begin{align}
H_0 &= \frac{v_F}{4 \pi} \int dx  \left[   \left( \partial_x \phi_R  \right)^2   +   \left( \partial_x \phi_L  \right)^2   \right] \ ,
\end{align}
where $\phi_{R/L}$ are bosonic fields describing the right/left moving modes traveling along the edge with velocity $v_F$. They satisfy a Kac-Moody commutation relation of the form $\left[ \phi_{R/L} (x), \phi_{R/L} (y) \right] = \pm i \pi \text{Sgn} (x-y)$. These fields are directly related to the quasiparticle creation and annihilation operators through the bosonization identity, $\psi_{R/L} (x,t) = \frac{U_{R/L}}{\sqrt{2 \pi a}} e^{\pm i k_F x} e^{-i \sqrt{\nu} \phi_{R/L} (x,t)}$, where $a$ is a short distance cut-off (typically the magnetic length), $k_F$ is the Fermi momentum, and $U_{R/L}$ are Klein factors. In particular, this identity gives us a direct connection between the bosonic fields and the quasiparticle density operator (and hence the current), as $\rho_{R/L} (x) = \pm e \frac{\sqrt{\nu}}{2 \pi} \partial_x \phi_{R/L} (x)$.

The Hall bar is further equipped with a quantum point contact, placed at position $x=0$, which allows tunneling between counter-propagating edges. In the weak backscattering regime, where quasiparticles are permitted to tunnel from one edge to the other through the bulk at the position of the QPC, this amounts to supplementing our Hamiltonian description with a tunneling term of the form
\begin{align}
H_\text{WB} &=  \Gamma_0  \psi_R^\dagger (0) \psi_L (0)  + \text{H.c.} \ ,
\label{eq:HWB}
\end{align}
where we introduced the bare tunneling constant $\Gamma_0$.

\emph{Deriving current and noise.---}
We are primarily interested in the fluctuations of the current flowing between the edge states: the backscattered current $I_B (t)$. The latter is readily obtained from the tunneling Hamiltonian, Eq.~\eqref{eq:HWB}, as the operator satisfying
\begin{align}
I_B (t) &= - e \dot{N}_R = i e^* \Gamma_0 \psi_R^\dagger (0,t) \psi_L (0,t) + \text{H.c.} \ ,
\end{align}
with the effective charge $e^* = \nu e$. Since no voltage bias is applied across the Hall bar, it is easy to convince oneself~\cite{Note1} that no current flows between the counter-propagating edge states, i.e. $\langle I_B (t) \rangle = 0$, independently of the respective temperature of the two edges.

Current-current correlations, however, do not vanish, as the finite temperature always lead to a nonzero contribution to the noise, through thermal fluctuations. More precisely, one can express the fluctuations of the backscattered current via the zero-frequency noise defined as
\begin{equation}
\cS_B = 2 \int_{-\infty}^{+\infty} d\tau \left[  \left\langle I_{B} (\tau) I_{B} (0) \right\rangle  - \left\langle I_{B} (\tau) \right\rangle \left\langle  I_{B} (0) \right\rangle  \right] \ .
\label{eq:noisedef}
\end{equation}
Using the Keldysh formalism, and relying on an expansion to second order in the tunneling Hamiltonian, the zero-frequency noise can be written as~\cite{Note1}
\begin{align}
\cS_B &= \left( \frac{e^* \Gamma_0}{\pi a} \right)^2  \int_{-\infty}^{+\infty} d\tau \exp \left[ \nu \cG_R \left( \tau \right) + \nu \cG_L \left( \tau \right) \right]  \ ,
\label{eq:zerofreqS}
\end{align}
where we introduced the bosonic Green's functions $\cG_{R/L} (\tau)$ typical of the chiral Luttinger liquid model of the FQHE edge states~\cite{martin05}.

Let us now specifically consider the situation of a temperature gradient between the two input ports of the QPC. There, the above expression for the noise does not lead to a tractable analytic form as a function of the two relevant temperatures. Instead, we resort to a perturbative expansion in the temperature difference, in the spirit of the scattering theory results \cite{lumbroso18}. 
Following the parametrization in temperature $T_{R/L} = \bT \pm \frac{\DT}{2}$, we then expand the result of Eq.~\eqref{eq:zerofreqS} in powers of $\DT$, noticing that, by symmetry under the exchange of reservoirs, the expansion contains no odd terms in the temperature difference. Up to fourth order in $\DT$, one has, after performing explicitly the resulting integrals
\begin{align}
\cS_B &= \cS_{\text{WB}}^{0} \left[ 1 + \cC_\nu^{(2)} \left( \frac{\DT}{2 \bT} \right)^2 + \cC_\nu^{(4)} \left( \frac{\DT}{2 \bT} \right)^4 \right]
\label{eq:noiseexpansion}
\end{align}
where $\cS_{\text{WB}}^{0}$ is the usual thermal noise in the weak backscattering limit [up to order $O(\Gamma_0^2)$] at temperature $\bT$, while $\cC_\nu^{(2)}$ and $\cC_\nu^{(4)}$ are numerical coefficients~\cite{Note1} depending only on the filling factor $\nu$. In particular, the leading order contribution has a prefactor
\begin{align} 
 \cC_\nu^{(2)} &=
 \nu
\left\{
 \frac{  \nu}{2\nu +1 }
\left[
\frac{\pi^2}{2}  - \psi' (\nu +1)
\right] -1
\right\} \ ,
\label{eq:calC2nu}
\end{align}
where $\psi' (x)$ is the first derivative of the digamma function. Similarly, a closed analytic expression for the coefficient $\cC_\nu^{(4)}$ is provided in the Supplemental Material~\cite{Note1}.

\begin{figure}[tb]
\includegraphics[width=8.6cm]{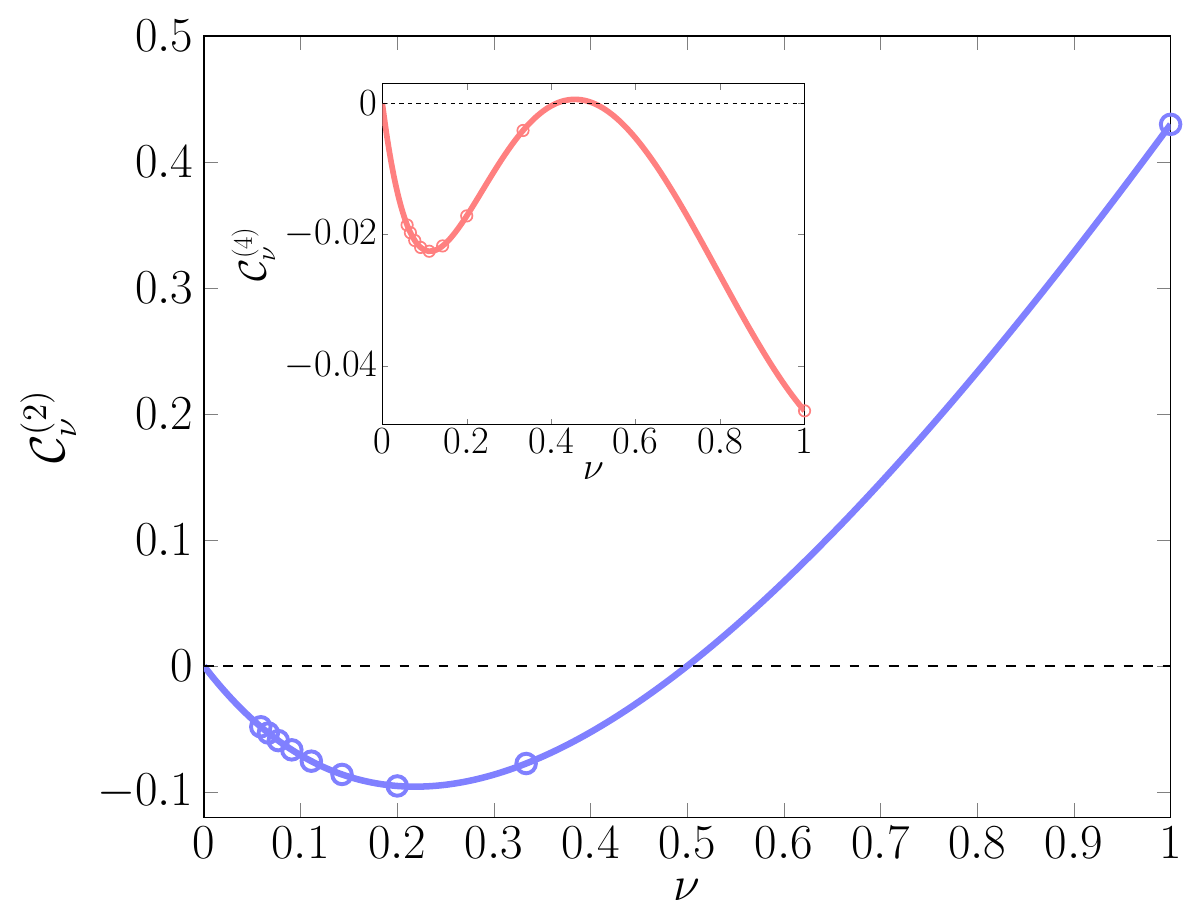}
\caption{Coefficients $\cC_\nu^{(2)}$ (main) and $\cC_\nu^{(4)}$ (inset) of the expansion of the noise in powers of the temperature difference [see Eq.~\eqref{eq:noiseexpansion}] as a function of the filling factor $\nu$. The circles correspond to the Laughlin fractions $\nu = 1/(2 n +1)$ for the first few values of $n= 0 , 1, \ldots, 8$.
}
\label{fig:CsvsNu}
\end{figure}

\emph{Main results.---}
As it happens, in the special situation of a Hall system at filling factor $\nu = 1$, our derivation yields $\cC_{\nu=1}^{(2)} = \left( \pi^2 - 6 \right)/9$, which matches with the non-interacting result obtained within scattering theory, and measured in metallic break junctions~\cite{lumbroso18}.

Our main results concern the fractional filling factors in the Laughlin sequence. We show in Fig.~\ref{fig:CsvsNu} the evolution of $\cC_\nu^{(2)}$ (main figure) and $\cC_\nu^{(4)}$ (inset) as a function of the filling factor $\nu$. From this, one readily sees that the second order term in the $\DT$-noise dominates over the fourth order one for all practical purposes, signaling that the expansion in $\DT$ is well controlled, despite the exotic power laws typical of the FQHE.

More importantly, our results presented in Fig.~\ref{fig:CsvsNu} suggest that  the delta-$T$ noise is a \emph{negative} contribution at leading order in $\DT$ for all filling factors in the Laughlin series. This represents a drastic difference, one of fundamental origin, with the non-interacting results where a positive correction was predicted and measured.  Indeed, such negative contribution implies a reduction of the noise, which constitutes an intriguing result as noise sources typically tend to add up. A decrease in fluctuations signals a key role from interactions, suggesting that such negative delta-$T$ noise contribution is a direct signature of the correlation effects characteristic of the edge states of the FQHE. 

Moreover, our results can be extended to the opposite regime of strong backscattering. This situation corresponds to a QPC near pinch-off, where the underlying Hall fluid is so depleted that only electrons can tunnel between the two halves of the Hall bar. We can repeat our derivation in this case~\cite{Note1} and show that it satisfies a duality transformation. The expression for the noise, Eq.~\eqref{eq:noiseexpansion}, is thus only slightly altered as
\begin{align}
\cS_B &= \cS_{\text{SB}}^{0} \left[ 1 + \cC_{1/\nu}^{(2)} \left( \frac{\DT}{2 \bT} \right)^2 + \cC_{1/\nu}^{(4)} \left( \frac{\DT}{2 \bT} \right)^4 \right]
\label{eq:noiseexpansionSB}
\end{align}
where $\cC_{1/\nu}^{(2)}$ and $\cC_{1/\nu}^{(4)}$ are readily obtained from the weak backscattering expressions upon substituting $\nu \to 1/\nu$, for fractional filling factors within the Laughlin sequence $\nu = 1/(2 n +1)$. In particular, this means that in this regime of electron tunneling at the QPC, one recovers a positive signal for the delta-$T$ noise contribution, like in the non-interacting case albeit with very different coefficients.

This is important in two ways. First, this makes a potential detection all the more easy from an experimental standpoint, as the delta-$T$ noise component flips its sign when one adjusts the transmission of the QPC between the weak and the strong backscattering regimes. Second, this result may have significant implications for the physics at play here. Our results suggest that negative delta-$T$ noise can be directly tied to the tunneling of quasiparticles, as the sign of this noise contribution reverts back to the non-interacting case when electrons tunnel through the QPC. Similarly, one can show following somewhat similar calculations~\cite{Note1} that weakly coupled non-chiral Luttinger liquids (e.g. coupled nanowires, or crossed nanotubes) cannot lead to negative backscattered noise when biased with a temperature difference. This underlines the importance of anyonic effects in the occurrence of negative delta-$T$ noise, which cannot be viewed as purely due to interactions. Instead, this effect has more to do with the interplay of strong interaction and the fractional statistics of the anyons exchanged at the QPC. Actually, the appearance of negative current correlations in other setups involving fractional edge states has been associated with anyonic statistics~\cite{rosenow16, lee19}. In Ref.~\onlinecite{lee19}, it was argued that negative excess shot noise was associated with so-called "topological vaccuum bubbles", an anyon process which has no counterpart for fermions. A similar mechanism involving the exchange of thermally excited anyons between the two edges at the QPC could be at play in the present device. Such a connection constitutes a fascinating incentive for future work in this direction.

Incidentally, in the (somewhat fictitious) situation of filling factor $\nu=1/2$, one can show that all coefficients $\cC_{\nu=1/2}^{(n)}$ reduce to zero, so that the delta-$T$ noise seems to exactly vanish in this case. This is actually an artifact of the weak backscattering limit. Using a refermionization approach~\cite{chamon96, sharma03}, one can account for the tunneling at the QPC at all orders in $\Gamma_0$, and show that the lowest order contribution to the delta-$T$ noise is actually of order $\Gamma_0^4$, falling beyond our present perturbative treatment.

\emph{Voltage dependence.---}
For completeness, we now consider the case where both a bias voltage and a temperature difference are applied across the Hall bar. Here, one can still expand in powers of the temperature difference, generalizing the expression obtained in Eq.~\eqref{eq:noiseexpansion} to account for the finite voltage. In particular, the prefactor is now given by  the  voltage-dependent noise $\cS_{\text{WB}}^{0} (V)$, which coincides with the expected backscattering noise of a QPC at temperature $\bT$ in the presence of a voltage bias V~\cite{kane94,chamon96}. Similarly, the coefficients of the expansion in $\DT$ must now be replaced with voltage-dependent coefficients $\cC_\nu^{(2)} (V)$ and $\cC_\nu^{(4)} (V)$.

\begin{figure}[tb]
\includegraphics[width=8.6cm]{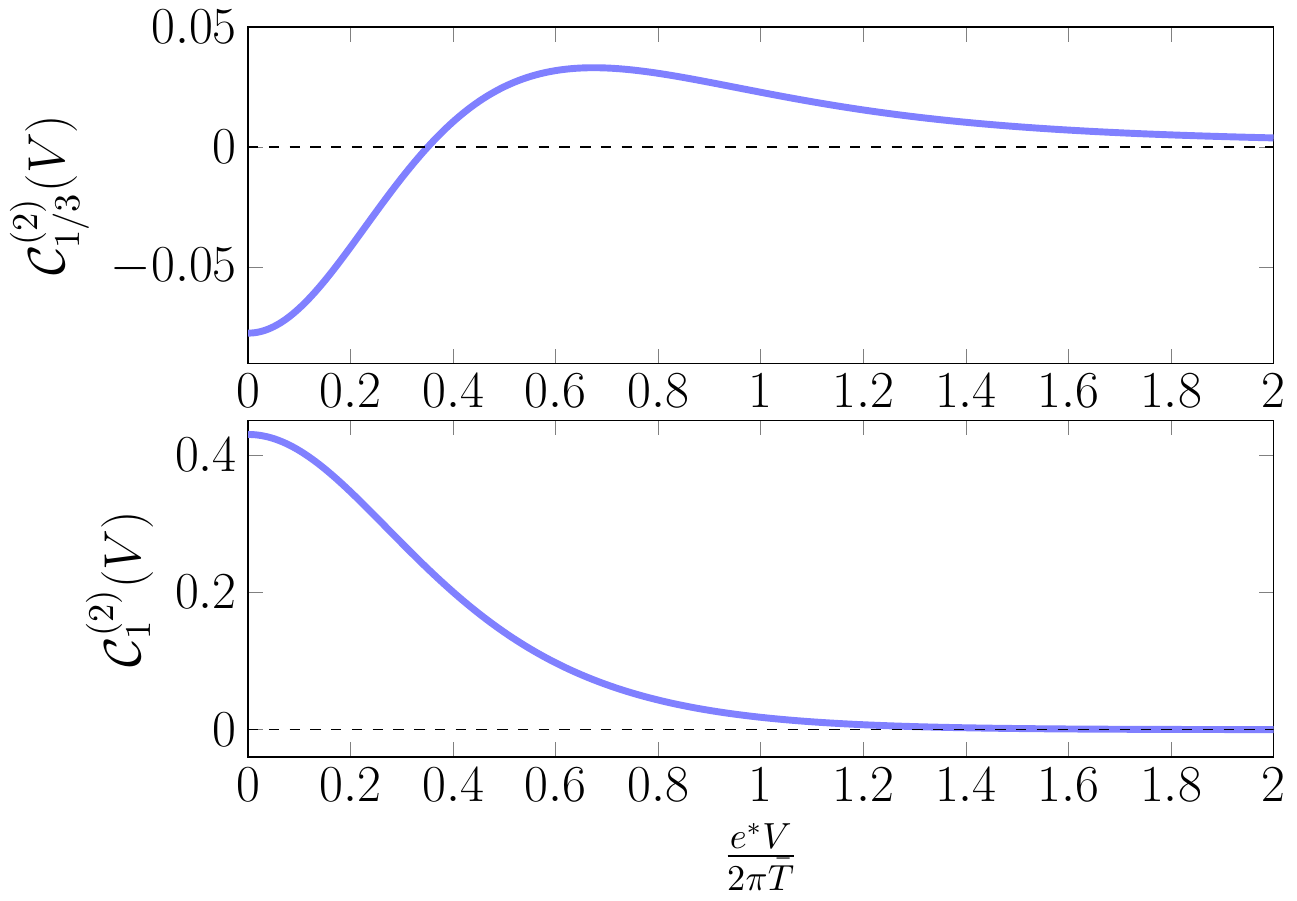}
\caption{Coefficients $\cC_{1/3}^{(2)} (V)$ and $\cC_1^{(2)} (V)$ of the expansion of the noise in powers of the temperature difference represented as a function of the applied bias across the Hall bar. 
}
\label{fig:Cnu2ofV}
\end{figure}

We show in Fig.~\ref{fig:Cnu2ofV} the evolution of the coefficients $\cC_{1/3}^{(2)} (V)$ and $\cC_1^{(2)} (V)$ of the leading contribution to the delta-$T$ noise, obtained as a function of voltage for filling factor $\nu=1/3$ and $\nu=1$ respectively. Not only do these results recover the previously obtained values in the limit of vanishingly small voltage (compare with Fig.~\ref{fig:CsvsNu}), they also predict that the coefficients of the expansion should tend toward zero for large enough voltage bias, as expected. Even more importantly, this shows that while in the non-interacting case the sign of $\cC_1^{(2)} (V)$ is fixed positive (as can be verified from the analytic expression~\cite{Note1}), in the fractional case, there exists a voltage scale where the coefficient $\cC_{1/3}^{(2)} (V)$ changes sign.

\emph{Experimental realization.---}
We now look more carefully at a potential experimental realization that could reveal our results concerning the delta-$T$ noise. The considered setup would be a 4-terminal device similar to the one schematically represented in Fig.~\ref{fig:setup}. The working principle of such a device would be to heat up contact 1, while leaving contact 2 at base temperature, and ensuring that no net current flows between edge states. 

Experimentally, there is no way to directly access the fluctuations of the backscattered current. Instead, the measurement is performed on contacts 3 and 4, allowing to probe the output currents $I_3$ and $I_4$ as well as the auto-correlations $\cS_{33}$ and $\cS_{44}$, and the cross-correlations $\cS_{34}$. The latter should be favored since cross-correlations of the currents vanish at equilibrium, thus increasing the visibility of the relevant signal. Following standard calculations~\cite{campagnano16}, one can show that the zero-frequency crossed correlations can be related to the backscattered noise~\cite{Note1}, allowing us to introduce the reduced noise $\tilde{\cS} = \cS_{34} - 4 \bT G_4 = - \cS_B$, where $G_4 = \frac{\partial I_4}{\partial V}$ is the differential conductance measured from contact 4.

In practice, the average temperature $\bT$ is not the most convenient quantity from an experimental standpoint. To circumvent this, it may be useful to slightly alter our parametrization, introducing instead $T_L = T_\text{cold}$ and $T_R = T_\text{cold} + \DT$, and measure the reduced noise in excess compared to the equilibrium situation. Rewriting our results with this prescription, we have, up to second order in $\DT$,
\begin{align}
\Delta \tilde{\cS} = \tilde{\cS} - \tilde{\cS}^\text{eq} &=  \tilde{\cS}^\text{eq} \left( 2 \nu - 1 \right) \frac{\DT}{2 T_\text{cold}} \nonumber \\
&\times \left[ 1 + \left( \nu - 1 + \frac{\cC_\nu^{(2)}}{2 \nu -1} \right) \frac{\DT}{2 T_\text{cold}}  \right] \ ,
\end{align}
where $\tilde{\cS}^\text{eq} =  - 4 T_\text{cold} G_4^0$ (as $\cS_{34}^\text{eq} = 0$) and $G_4^0$ is the differential conductance in the absence of a temperature bias. This allows to extract the coefficient $\cC_\nu^{(2)}$ directly from the experimental measurement of cross-correlations and conductance, with and without the temperature gradient.

In all generality, delta-$T$ noise may be partly eclipsed by thermoelectric effects. Indeed, as a result of a temperature difference, an electric current may develop even in the absence of any applied voltage, a phenomenon known as the Seebeck effect~\cite{tritt06}. This thermally induced current can, in turn, lead to conventional shot noise. Following the results displayed in Fig.~\ref{fig:Cnu2ofV}, for such thermoelectric effects to overshadow the negative contribution in $\DT$, one would need to generate a rather large voltage difference $V$, satisfying $e^* V \gtrsim 2.2 \bT$. This is, however, unlikely to happen in quantum Hall devices. Not only the bulk is incompressible, and therefore expected not to contribute substantially to any relevant thermoelectric properties, but also the edge states are chiral, and can only overlap at a single point-like QPC, thus preserving particle-hole symmetry. Moreover, no equilibration is expected to take place over small enough distances (a few microns in relevant experiments) and the heat leakage into the bulk is similarly negligible for short enough edges, thus ruling out the impact of thermoelectric effects in realistic setups. Finally, it is also worth pointing out that in recent experiments on metallic break junctions~\cite{lumbroso18}, the thermoelectric contribution was shown to be orders of magnitude smaller than the delta-$T$ noise.

\emph{Conclusions.---}We have studied the peculiar signatures of a temperature difference across a QPC, showing that it leads to a negative contribution to the noise associated with the tunneling of Laughlin quasiparticles. We have characterized this so-called delta-$T$ noise in the weak and strong backscattering regimes and investigated its dependence on voltage bias. We argued that it is readily accessible in current experiments involving fractional edge states.

This work offers many interesting perspectives, accounting more carefully for equilibration (important for long edges) or thermoelectric effects (e.g. associated with an extended QPC~\cite{vannucci15}). The most intriguing one concerns the extension to more exotic states, beyond the Laughlin filling factors, where many new fascinating effects should occur as a consequence of the more complex structure of the edge states (and in particular, the presence of neutral modes~\cite{ferraro10,shtanko14}). 

\begin{acknowledgments}
We are grateful to  L. Raymond, D.C. Glattli, G. F\`eve, H. Bartolomei, and P. Degiovanni for useful discussions. 
The project leading to this publication has received funding from Excellence Initiative of  Aix-Marseille University - A$^*$MIDEX, a French "Investissements d'Avenir" program (AMX-19-IET-009).
\end{acknowledgments}

\bibliography{DeltaT_biblio}


\onecolumngrid
\newpage

\appendix

\section*{SUPPLEMENTAL MATERIAL}

\section{Non-interacting fermions: a scattering theory calculation}

We consider here a two-terminal device where two fermionic non-interacting spinless reservoirs are brought together, thereby defining a junction where we assume a single conduction channel for simplicity. As is common in such devices, the tunnel barrier between the reservoirs is described by a transmission coefficient $\cT(E)$ which we approximate by assuming it is energy-independent, $\cT(E) \simeq \cT$. The reservoirs, labeled $i=1$ and 2 respectively, are characterized by their chemical potential $\mu_i$ and temperature $T_i$, leading to a description in terms of the Fermi distribution $f_i (E) = \left[ 1 + e^\frac{E - \mu_i}{T_i}  \right]^{-1}$.

Using scattering theory, the current through the junction is readily obtained from
\begin{align}
\langle \hat{I} \rangle = \frac{e}{2 \pi} \int dE~ \cT \left[ f_1 (E) - f_2 (E) \right]
\end{align}
In the absence of a potential bias, the chemical potential of the two reservoirs are equal, $\mu_1 = \mu_2 = \mu$, and the resulting energy integral for the current identically vanishes, independently of the temperature on both sides of the junction.

Fluctuations away from this average value are characterized by the noise, i.e. the current-current correlations, which we consider here at zero-frequency. Within the scattering theory formalism, one obtains the standard expression for the zero-frequency noise
\begin{align}
S &= 2 \int dt \left[ \langle \hat{I} (t) \hat{I} (0) \rangle -  \langle \hat{I} (t) \rangle \langle \hat{I} (0) \rangle \right] \nonumber \\
&= \frac{e^2}{\pi} \int dE \left\{ \cT \left[ f_1 (E) \left( 1 - f_1 (E) \right) + f_2 (E) \left( 1 - f_2 (E) \right) \right] + \cT \left( 1- \cT \right) \left( f_1(E) - f_2(E) \right)^2  \right\}
\end{align}
The first term corresponds to thermal-like noise, while the last one is a non-equilibrium contribution. 

We now focus on the specific situation where no bias is applied, $\mu_1 = \mu_2 = \mu$, but the temperature of the two reservoirs are different and parametrized by
\begin{align}
T_1 & = \bT - \frac{\DT}{2} \\
T_2 & = \bT + \frac{\DT}{2} 
\end{align}

Noticing that $  f_i (E) \left( 1 - f_i (E) \right) =  - T_i \frac{d f_i (E)}{dE}$, the first contribution is readily obtained and reduces to
\begin{align}
\frac{e^2}{\pi} \cT \int dE \left[ f_1 (E) \left( 1 - f_1 (E) \right) + f_2 (E) \left( 1 - f_2 (E) \right) \right] = \frac{e^2}{\pi} \cT \left[ T_1+ T_2 \right] = 2 \frac{e^2}{\pi} \cT \bT
\end{align}

The remaining, non-equilibrium, contribution does not reduce to a simple analytic form. Instead, assuming a small temperature difference compared with the average temperature of the reservoirs, we rely on an expansion in powers of the parameter $\frac{\DT}{2 \bT}$. Indeed, one has
\begin{align}
 \left[ f_1(E) - f_2(E) \right]^2 = \left( \frac{\DT}{2 \bT} \right)^2 \left( 2 \bT \left. \frac{\partial f(E)}{\partial T} \right|_{T = \bT} \right)^2 + \frac{4}{3} \left( \frac{\DT}{2 \bT} \right)^4  \bT \left. \frac{\partial f(E)}{\partial T} \right|_{T = \bT} \bT^3 \left. \frac{\partial^3 f(E)}{\partial T^3} \right|_{T = \bT} + O \left[ \left( \frac{\DT}{2 \bT} \right)^6 \right]
\end{align}
where $f(E) = \left[ 1 + e^\frac{E - \mu}{T}  \right]^{-1}$.

The resulting integrals can be carried out as
\begin{align}
\int dE \left( 2 \bT \left. \frac{\partial f(E)}{\partial T} \right|_{T = \bT} \right)^2 &= \bT \int du ~\frac{ 4 u^2 e^{2 u}}{\left( 1+ e^u \right)^4} \nonumber \\
&= 2 \bT \frac{\pi^2 - 6}{9}
\end{align}
and 
\begin{align}
\int dE~ \bT \left. \frac{\partial f(E)}{\partial T} \right|_{T = \bT} \bT^3 \left. \frac{\partial^3 f(E)}{\partial T^3} \right|_{T = \bT}  &=  \bT \int du ~u^2 e^{2 u}
\frac{ 6 + 6 u + u^2 - 4 e^u \left( u^2 -3 \right) + e^{2 u } \left(6 -6 u + u^2 \right)}{\left( 1+ e^u \right)^6} \nonumber \\
&= - \bT \frac{7 \pi^4 - 75 \pi^2 + 90}{450}
\end{align}
where we introduced the reduced variable $u = \frac{E - \mu}{ \bT}$.

Putting all the contributions back together, one is left with
\begin{align}
S \approx 2 \frac{e^2}{\pi}  &\left\{ \cT \bT + \cT \left( 1- \cT \right) \bT \left[ \frac{\pi^2 - 6}{9} \left( \frac{\DT}{2 \bT} \right)^2 -\frac{7 \pi^4 - 75 \pi^2 + 90}{675} \left( \frac{\DT}{2 \bT} \right)^4 \right] \right\}
\end{align}


\section{Backscattered current and noise}

The system considered here is a Hall bar, in the fractional quantum Hall regime, restricting ourselves to filling factors in the Laughlin sequence, i.e. $\nu = 1/(2 n +1)$. The Hall bar is equipped with a quantum point contact (QPC), placed at position $x=0$. In the weak backscattering regime, quasiparticles are allowed to tunnel from one edge to the other through the bulk at the position of the QPC, leading to a tunneling Hamiltonian of the form
\begin{align}
H_\text{WB} &= \Gamma_0  e^{ i e^* V t}  \psi_R^\dagger (0) \psi_L (0) +  \text{H.c.}
\label{eq:HWB}
\end{align}
where we introduced the effective charge $e^* = \nu e$ and used the Peierls substitution to make the voltage appear explicitly in the tunneling Hamiltonian rather than in the contacts.

The backscattered current is readily obtained from this tunneling Hamiltonian and reads
\begin{align}
I_B (t) &= - e \dot{N}_R =  i e \left[ N_R ,  H_\text{WB} \right] \nonumber \\
&= i e^*  \left[ \Gamma_0 e^{ i e^* V t} \psi_R^\dagger (0,t) \psi_L (0,t)   -   \Gamma_0 e^{ -i e^* V t} \psi_L^\dagger (0,t) \psi_R (0,t)  \right]
\end{align}

The expectation value of the backscattered current can be conveniently expressed in the Keldysh formalism as
\begin{align}
\cI_B = \langle I_B (t) \rangle &= \frac{1}{2} \sum_{\eta = +,-}  \left\langle T_K I_B \left( t^{(\eta)} \right) e^{- i \int_{{\cal C}} dt' H_T (t')}  \right\rangle
\end{align}
which is further expanded up to second order in the tunneling parameter as (first order perturbation in $H_T$)
\begin{align}
\cI_B 
&= \frac{e^*}{2} \Gamma_0^2 \int dt' \sum_{\eta, \eta'} \eta'  
\left\{ e^{ i e^* V (t-t')}
  \left\langle T_K 
  \psi_R^\dagger \left(0,t^{(\eta)} \right) \psi_R \left(0,t'^{(\eta')} \right)
 \right\rangle \left\langle T_K
\psi_L \left(0,t^{(\eta)} \right)   \psi_L^\dagger \left(0,t'^{(\eta')} \right)   
\right\rangle  \right. \nonumber \\
& \qquad \qquad -
\left.  e^{ -i e^* V (t-t')}
 \left\langle T_K 
 \psi_L^\dagger \left(0,t^{(\eta)} \right) \psi_L \left(0,t'^{(\eta')} \right) 
 \right\rangle \left\langle T_K 
 \psi_R \left(0,t^{(\eta)} \right) \psi_R^\dagger \left(0,t'^{(\eta')} \right)  
\right\rangle  \right\}
\nonumber \\
&= -2 i \frac{e^* \Gamma_0^2}{\left(2 \pi a \right)^2}  \int_{-\infty}^\infty d \tau \sin \left( e^* V \tau \right)  e^{\nu \cG_R \left( -\tau \right)} e^{\nu \cG_L \left( -\tau \right)}  
\label{eq:avgIB}
\end{align}
where we used that
\begin{equation}
 \left\langle T_K \psi_\mu^\dagger (0,t^{(\eta)}) \psi_\mu (0,t'^{(\eta')}) \right\rangle = \left\langle T_K \psi_\mu (0,t^{(\eta)}) \psi_\mu^\dagger (0,t'^{(\eta')}) \right\rangle = \frac{1}{2 \pi a} \exp \left[ \nu \cG_\mu \left( \sigma_{\eta \eta'} (t-t') \right) \right]
\end{equation}
and introduced $\sigma_{\eta \eta'} (t-t') = \eta' \left[ (1-\delta_{\eta \eta'}) (t-t') + \delta_{\eta \eta'} |t-t'| \right]$.

The bosonic Green's function is given at finite temperature by
\begin{equation}
\cG_\mu (\tau) = - \log \left[ \frac{\sinh \left( \frac{\pi}{\beta_\mu} (i \tau_0 - \tau) \right) }{ \sinh \left( i \frac{\pi}{\beta_\mu} \tau_0 \right) } \right]
\label{eq:bosonicGF}
\end{equation}
where $\beta_{\mu} = 1/T_\mu$ is the inverse temperature of the considered lead (recall that $k_B=1$) and $\tau_0 = a/v_F$ is a short-time (or high-energy) cutoff.

The current noise can be written in terms of the backscattered current at the QPC as
\begin{equation}
S_B (t,t') = \left\langle I_{B} (t) I_{B} (t') \right\rangle - \left\langle I_{B} (t) \right\rangle \left\langle I_{B} (t') \right\rangle
\label{eq:noisedef}
\end{equation}
It is then similarly obtained through perturbative expansion in the tunnel Hamiltonian writing, to second order in $\Gamma_0$
\begin{align}
S_B (t,t') &= \left\langle T_K \Delta I_{B} (t^{(+)}) \Delta I_{B} (t'^{(-)}) \exp\left(  - i \sum_{\eta = \pm} \eta \int_{-\infty}^{+\infty} dt'' H_T (t''^{(\eta)}) \right)  \right\rangle \nonumber \\
&= \left( e^* \Gamma_0 \right)^2 \left[ e^{ i e^* V (t-t') }  \left\langle T_K \psi_R^\dagger \left(0,t^{(+)} \right) \psi_R \left(0,t'^{(-)} \right) \right\rangle  \left\langle T_K \psi_L \left(0,t^{(+)} \right) \psi_L^\dagger \left(0,t'^{(-)} \right) \right\rangle \right.  \nonumber \\
& \qquad  \qquad + \left. e^{ - i e^* V (t-t') }  \left\langle T_K \psi_L^\dagger \left(0,t^{(+)} \right) \psi_L \left(0,t'^{(-)} \right) \right\rangle  \left\langle T_K \psi_R \left(0,t^{(+)} \right) \psi_R^\dagger \left(0,t'^{(-)} \right) \right\rangle \right] \nonumber \\
&= 2 \left( \frac{e^* \Gamma_0}{2 \pi a} \right)^2 \cos \left[ e^* V (t-t') \right] \exp \left[ \nu \cG_R \left( t'-t \right) + \nu \cG_L \left( t'-t \right) \right]
\end{align}
From this, one readily sees that the noise only depends on the time difference, so that $S_B ( t, t') = S_B (t-t')$. One can thus define the zero-frequency noise as
\begin{align}
\cS_B &= 2 \int d\tau ~ S_B \left( \tau \right) \nonumber \\
&= \left( \frac{e^* \Gamma_0}{\pi a} \right)^2  \int d\tau \cos \left(  e^* V \tau \right) \exp \left[ \nu \cG_R \left( - \tau \right) + \nu \cG_L \left( - \tau \right) \right]  \nonumber \\
&= \left( \frac{e^* \Gamma_0}{\pi a} \right)^2  \int d\tau \cos \left(  e^* V \tau \right)   \left[ \frac{ \sinh \left( i \pi T_R \tau_0 \right) }{\sinh \left( \pi T_R (i \tau_0 +\tau) \right) } \right]^{\nu} \left[ \frac{ \sinh \left( i \pi T_L \tau_0 \right) }{\sinh \left( \pi T_L (i \tau_0 +\tau) \right) } \right]^{\nu}
\label{eq:zerofreqS}
\end{align}

\section{Temperature biased case: $T_R \neq T_L$ and $V = 0$}

Let us first consider the situation of zero bias voltage, and focus on the effect of a temperature difference between the two input ports of the QPC. In this particular situation, the expression for the backscattered current becomes trivial: as one can readily see from the general expression given in Eq.~\eqref{eq:avgIB}, the current vanishes in the absence of a voltage bias, no matter what the temperature difference is.

Things are different for the noise, as the temperature difference induces extra fluctuations of the current compared to the equilibrium case, which are then susceptible to be partitioned at the QPC. While a fully analytic expression is out of reach, we instead resort to a perturbative expansion in the temperature difference.

Introducing the notations
\begin{align}
T_{R/L} = \bT \pm \frac{\DT}{2}
\end{align}
and expanding the result of Eq.~\eqref{eq:zerofreqS} to fourth order in $\DT$, one has
\begin{align}
\cS_B &= \cS_0 + \left( \frac{\DT}{2 \bT} \right)^2 \cS_2 + \left( \frac{\DT}{2 \bT} \right)^4 \cS_4 + O \left[  \left( \frac{\DT}{2 \bT} \right)^6 \right]
\end{align}
Notice that, by symmetry of the parametrization in temperature, there are no linear terms in the temperature difference.

The various contributions can then be computed separately. The leading-order term corresponds to the thermal noise, it is given by
\begin{align}
\cS_0 & = \left( \frac{e^* \Gamma_0}{\pi a} \right)^2  \int d\tau \left[ \frac{ \sinh \left( i \pi \bT \tau_0 \right) }{\sinh \left( \pi \bT (i \tau_0 +\tau) \right) } \right]^{2 \nu} 
 = \frac{ e^2}{2 \pi} \left( \frac{ 2 \nu \Gamma_0}{v_F} \right)^2  \bT \left(2  \pi \bT \tau_0 \right)^{2 \nu-2} 
 \frac{ \Gamma \left(\nu  \right)^2 }{\Gamma \left( 2\nu \right)}
 \label{eq:exprS0}
\end{align}

The leading correction reads
\begin{align}
\cS_2 & = \left( \frac{e^* \Gamma_0}{\pi a} \right)^2  \int d\tau \left[ \frac{ \sinh \left( i \pi \bT \tau_0 \right) }{\sinh \left( \pi \bT (i \tau_0 +\tau) \right) } \right]^{2 \nu} 
\left[ 
\frac{ \nu \pi^2 \bT^2 \tau_0^2}{\sinh^2 \left( i \pi \bT \tau_0 \right)}
+
\frac{ \nu \pi^2 \bT^2 (i \tau_0 +\tau)^2}{\sinh^2 \left( \pi \bT (i \tau_0 +\tau) \right)}
\right] \nonumber \\
 & = 
\frac{ e^2}{2 \pi}  \left( \frac{ 2 \nu \Gamma_0}{v_F} \right)^2
\left(2 \pi \bT \tau_0 \right)^{2 \nu-2} \nu \bT \frac{ \Gamma \left(\nu  \right)^2 }{\Gamma \left( 2\nu \right)} 
\left\{
 \frac{  \nu}{2\nu +1 }
\left[
\frac{\pi^2}{2}  - \psi' (\nu +1)
\right] -1
\right\}  
\end{align}
and the next order one is yet more involved, writing
\begin{align}
\cS_4  &= \left( \frac{e^* \Gamma_0}{\pi a} \right)^2  \int d\tau \left[ \frac{ \sinh \left( i \pi \bT \tau_0 \right) }{\sinh \left( \pi \bT (i \tau_0 +\tau) \right) } \right]^{2 \nu} 
\left[ 
\frac{\nu (\nu-1)}{2} \frac{\pi^4 \bT^4 \tau_0^4}{\sinh^4 \left( i \pi \bT \tau_0 \right)}
+
\frac{\nu}{3} \frac{ \pi^4 \bT^4 (i \tau_0 +\tau)^4}{\sinh^2 \left( \pi \bT (i \tau_0 +\tau) \right)}
\right. \nonumber \\
& \qquad \qquad  \qquad \qquad  \qquad \qquad  \qquad \qquad  \qquad  - \frac{\nu}{3}
\frac{ \pi^4 \bT^4 \tau_0^4}{\sinh^2 \left( i \pi \bT \tau_0 \right)}
+
\frac{\nu (\nu+1)}{2} \frac{\pi^4 \bT^4 (i \tau_0 +\tau)^4}{\sinh^4 \left( \pi \bT (i \tau_0 +\tau) \right)}
 \nonumber \\
&  \qquad \qquad  \qquad \qquad  \qquad \qquad  \qquad \qquad  \qquad  +
\left.
\nu^2 
\frac{ \pi^2 \bT^2 \tau_0^2}{\sinh^2 \left( i \pi \bT \tau_0 \right)}
\frac{ \pi^2 \bT^2 (i \tau_0 +\tau)^2}{\sinh^2 \left( \pi \bT (i \tau_0 +\tau) \right)}
\right] \nonumber \\
 & =  \frac{ e^2}{2 \pi}   \left( \frac{ 2 \nu \Gamma_0}{v_F} \right)^2  \bT \left(2  \pi \bT \tau_0 \right)^{2 \nu-2} 
 \frac{ \Gamma \left(\nu  \right)^2 }{\Gamma \left( 2\nu \right)} \nonumber \\
&  \qquad \qquad \times \left\{
 \nu \frac{\pi^4 \nu^2 \left( 4 + 3 \nu \right) - 12 \pi^2 \nu \left( 2 \nu^2 + 3 \nu -3 \right) +12 \left( 4 \nu^3 +4 \nu^2 -5 \nu -3 \right)}{24 \left( 4 \nu^2 +8 \nu +3 \right)}
\right. \nonumber \\
&  \qquad \qquad \qquad 
 + \nu^2 \frac{4 \nu^2 + 6 \nu - 6 - \pi^2 \nu \left( 4 + 3 \nu \right)}{8 \nu^2 +16 \nu + 6} \psi'  \left(\nu +1 \right)
  + \nu^3 \frac{4+3\nu}{2 \left( 4 \nu^2 + 8 \nu + 3\right)} \left[ \psi'  \left(\nu +1 \right) \right]^2 \nonumber \\
&  \qquad \qquad \qquad \left.
  + \nu^3 \frac{4+3\nu}{12 \left( 4 \nu^2 + 8 \nu + 3\right)} \psi^{(3)}  \left(\nu +1 \right)
 \right\}
\end{align}

Bringing all these contributions together, one is left with the following expression for the noise
\begin{align}
\cS_B &=
 \cS_\text{WB}^0 \left[ 1 + \left( \frac{\DT}{2 \bT} \right)^2 \cC_\nu^{(2)} + \left( \frac{\DT}{2 \bT} \right)^4 \cC_\nu^{(4)} \right]
\label{eq:fullcalS}
\end{align}
where $ \cS_\text{WB}^0 =  \cS_0$ and the coefficients $\cC_\nu^{(n)}$ take the form
\begin{align} 
\label{eq:calC2nu}
 \cC_\nu^{(2)} &=
 \nu
\left\{
 \frac{  \nu}{2\nu +1 }
\left[
\frac{\pi^2}{2}  - \psi' (\nu +1)
\right] -1
\right\} \\
\cC_\nu^{(4)} &=
 \nu \frac{\pi^4 \nu^2 \left( 4 + 3 \nu \right) - 12 \pi^2 \nu \left( 2 \nu^2 + 3 \nu -3 \right) +12 \left( 4 \nu^3 +4 \nu^2 -5 \nu -3 \right)}{24 \left( 4 \nu^2 +8 \nu +3 \right)}
 \nonumber \\
& \quad 
 + \nu^2 \frac{4 \nu^2 + 6 \nu - 6 - \pi^2 \nu \left( 4 + 3 \nu \right)}{8 \nu^2 +16 \nu + 6} \psi'  \left(\nu +1 \right)
  + \nu^3 \frac{4+3\nu}{2 \left( 4 \nu^2 + 8 \nu + 3\right)} \left[ \psi'  \left(\nu +1 \right) \right]^2 \nonumber \\
& \quad 
  + \nu^3 \frac{4+3\nu}{12 \left( 4 \nu^2 + 8 \nu + 3\right)} \psi^{(3)}  \left(\nu +1 \right)
\end{align}

As a first step, one can check that in the special situation of filling factor $\nu = 1$, one recovers the expected values for the prefactors above, namely
\begin{align} 
 \cC_1^{(2)} &= \frac{\pi^2}{9} - \frac{2}{3} \simeq 0.42996\\
\cC_1^{(4)} &= -\frac{7 \pi^4}{675} + \frac{\pi^2}{9} - \frac{2}{15} \simeq - 0.04688
\end{align}
where we used that $\psi' \left( 2 \right) = \frac{\pi^2}{6} - 1$ and $\psi^{(3)} \left( 2 \right) = \frac{\pi^4}{15} - 6 $.

Quite surprisingly, using some specific values of the digamma function and its derivatives, one can also show that for a (fictitious) filling factor of $\nu = 1/2$, one has for the coefficients of the expansion
\begin{align} 
 \cC_{1/2}^{(2)} &=  \frac{1}{2}
\left\{
 \frac{1}{4 }
\left[
\frac{\pi^2}{2}  - \psi' \left( \frac{3}{2} + 1 \right)
\right] -1
\right\} = 0 \\
\cC_{1/2}^{(4)} &=   \frac{11 \pi^4   + 48 \pi^2  - 384}{3072}
 -  \frac{8 + 11 \pi^2 }{256} \psi'  \left(\frac{1}{2} +1 \right)
  + \frac{11}{256} \left[ \psi'  \left(\frac{1}{2} +1 \right) \right]^2 
  + \frac{11}{1536} \psi^{(3)}  \left(\frac{1}{2} +1 \right) =0
\end{align}
where we used that $\psi' \left( \frac{3}{2} \right) = \frac{\pi^2}{2} - 4$ and $\psi^{(3)} \left( \frac{3}{2} \right) = \pi^4 - 96 $.  These  results raise the question of the fate of the $\DT$ noise in the special situation of filling factor $\nu=1/2$. Having a closer look at the full expression for the noise, we have for this special filling factor
\begin{align}
\cS_B & = \left( \frac{e^* \Gamma_0}{\pi a} \right)^2  \int d\tau \sqrt{ \frac{ \sinh \left( i \pi T_R \tau_0 \right) }{\sinh \left( \pi T_R (i \tau_0 +\tau) \right) }}  \sqrt{ \frac{ \sinh \left( i \pi T_L \tau_0 \right) }{\sinh \left( \pi T_L (i \tau_0 +\tau) \right) }}
\end{align}
As it turns out, in the limit of vanishingly small cutoff $a \to 0$, this result is independent of $\DT$, so that the delta-$T$ noise exactly vanishes at this order in the tunneling parameter $\Gamma_0$.

\section{Strong backscattering regime}

All the results presented so far have been obtained in the regime of weak backscattering. It is, however, interesting to investigate the fate of the $\DT$ noise in the opposite regime of strong backscattering, where the tunneling Hamiltonian now reads
\begin{align}
H_\text{SB} &=  \Gamma  e^{ i e V t}  \Psi_R^\dagger (0) \Psi_L (0) + \text{H.c.}
\end{align}
where $\Psi_{R/L}$ is the operator associated with the annihilation of a full electron.
 
This leads to substantial modifications in the expressions for the average current and zero-frequency noise, which now become
\begin{align}
\cI_B &=  -2 i \frac{e \Gamma}{\left(2 \pi a \right)^2}  \int d \tau~  \sin \left(  e V \tau \right)\exp \left[ \frac{\cG_R \left( - \tau \right) + \cG_L \left( - \tau \right)}{\nu} \right]    \\
\cS_B & = \left( \frac{e \Gamma}{\pi a} \right)^2  \int d\tau~ \cos \left(  e V \tau \right) \exp \left[ \frac{\cG_R \left( - \tau \right) + \cG_L \left( - \tau \right)}{\nu} \right]  
\end{align}
These expressions could be readily obtained from the ones derived in the weak backscattering regime upon performing a duality transformation, i.e. substituting $e^* \to e$, $\Gamma_0 \to \Gamma$ and $\nu \to \frac{1}{\nu}$.

It follows that one can similarly extend our results for the $\DT$ noise to this regime of strong backscattering, leading to
\begin{align}
\cS_B &=  \cS_\text{SB}^0 \left[ 1 + \left( \frac{\DT}{2 \bT} \right)^2 \cC_{1/\nu}^{(2)} + \left( \frac{\DT}{2 \bT} \right)^4 \cC_{1/\nu}^{(4)} \right]
\end{align}
where one has
\begin{align}
 \cS_\text{SB}^0 & = \frac{ e^2}{2 \pi} \left( \frac{ 2 \Gamma}{v_F} \right)^2  \bT \left(2  \pi \bT \tau_0 \right)^{\frac{2}{\nu}-2} 
 \frac{ \Gamma \left( \frac{1}{\nu}  \right)^2 }{\Gamma \left( \frac{2}{\nu} \right)}
\end{align}
and the coefficients $\cC_{1/\nu}^{(n)}$ take the form
\begin{align} 
 \cC_{1/\nu}^{(2)} &=
 \frac{1}{\nu}
\left\{
 \frac{1}{2 + \nu }
\left[
\frac{\pi^2}{2}  - \psi' \left(\frac{1}{\nu} +1 \right)
\right] -1
\right\} \\
\cC_{1/\nu}^{(4)} &=
 \frac{1}{\nu^2} \frac{\pi^4  \left( 4 \nu + 3 \right) - 12 \pi^2  \left( 2  + 3 \nu -3 \nu^2 \right) +12 \left( 4  +4 \nu -5 \nu^2 -3 \nu^3 \right)}{24 \left( 4 +8 \nu +3 \nu^2 \right)}
 \nonumber \\
& \quad 
 + \frac{1}{\nu^2} \frac{4  + 6 \nu - 6 - \pi^2  \left( 4 \nu + 3  \right)}{8  +16 \nu + 6 \nu^2} \psi'  \left(\frac{1}{\nu} +1 \right)
  + \frac{1}{\nu^2} \frac{4 \nu +3}{2 \left( 4  + 8 \nu + 3 \nu^2 \right)} \left[ \psi'  \left(\frac{1}{\nu} +1 \right) \right]^2 \nonumber \\
& \quad 
  + \frac{1}{\nu^2} \frac{4 \nu + 3}{12 \left( 4 + 8 \nu + 3 \nu^2 \right)} \psi^{(3)}  \left(\frac{1}{\nu} +1 \right)
\end{align}
Note that, one readily sees from these results that the case $\nu = 1/2$ is no longer special, suggesting that the vanishing of the $\DT$ noise for this specific value of the filling factor has to do with the weak backscattering regime and does not extend beyond it.

\section{Voltage dependence}

We now extend the results of the previous sections, applying an external bias voltage in addition to the small temperature difference. Starting from the general expression of Eq.~\eqref{eq:zerofreqS} in the weak backscattering regime, and expanding in powers of the temperature difference $\DT$, we have, up to second order
\begin{align}
\cS_B &=  \left( \frac{e^* \Gamma_0}{\pi a} \right)^2 \int d \tau \cos \left( e^* V \tau \right) 
\left[ \frac{ \sinh \left( i \pi \bT \tau_0 \right) }{\sinh \left( \pi \bT (i \tau_0 +\tau) \right) } \right]^{2 \nu} 
\left\{ 1 
+ \left( \frac{\DT}{2 \bT} \right)^2 \left[ 
\frac{ \nu \pi^2 \bT^2 \tau_0^2}{\sinh^2 \left( i \pi \bT \tau_0 \right)}
+
\frac{ \nu \pi^2 \bT^2 (i \tau_0 +\tau)^2}{\sinh^2 \left( \pi \bT (i \tau_0 +\tau) \right)}
\right]
 \right. \nonumber \\
& \qquad  \qquad + \left( \frac{\DT}{2 \bT} \right)^4
\left[ 
\frac{\nu (\nu-1)}{2} \frac{\pi^4 \bT^4 \tau_0^4}{\sinh^4 \left( i \pi \bT \tau_0 \right)}
+
\frac{\nu}{3} \frac{ \pi^4 \bT^4 (i \tau_0 +\tau)^4}{\sinh^2 \left( \pi \bT (i \tau_0 +\tau) \right)}
 - 
 \frac{\nu}{3} \frac{ \pi^4 \bT^4 \tau_0^4}{\sinh^2 \left( i \pi \bT \tau_0 \right)}
\right. \nonumber \\
& \qquad \qquad  \qquad \qquad \qquad  \left.   \left.
+
\frac{\nu (\nu+1)}{2} \frac{\pi^4 \bT^4 (i \tau_0 +\tau)^4}{\sinh^4 \left( \pi \bT (i \tau_0 +\tau) \right)}
+
\nu^2 
\frac{ \pi^2 \bT^2 \tau_0^2}{\sinh^2 \left( i \pi \bT \tau_0 \right)}
\frac{ \pi^2 \bT^2 (i \tau_0 +\tau)^2}{\sinh^2 \left( \pi \bT (i \tau_0 +\tau) \right)}
\right]
\right\}
 \nonumber \\
\end{align}

This can be written in a similar form as before, only now involving voltage-dependent coefficients, as
\begin{align}
\cS_B &=
 \cS_\text{WB}^0 (V) \left\{ 1 + \left( \frac{\DT}{2 \bT} \right)^2 \cC_\nu^{(2)} (V) + O\left[ \left( \frac{\DT}{2 \bT} \right)^4 \right] \right\}
\end{align}
where the noise in the absence of a voltage difference reads
\begin{align}
\cS_\text{WB}^0 (V)  
&= \frac{e^2}{2 \pi}  \left( \frac{2 \nu \Gamma_0}{ v_F} \right)^2  \bT \left( 2 \pi \bT \tau_0 \right)^{2 \nu-2} 
\frac{ \left| \Gamma \left(\nu +i \frac{e^* V}{2 \pi \bT} \right)   \right|^2 }{\Gamma \left( 2\nu \right)}  \cosh \left( \frac{e^* V}{2 \bT} \right)
\end{align}
and the coefficient $\cC_\nu^{(2)} (V)$ is given by
\begin{align}
 \cC_\nu^{(2)} (V) &=
 - \nu + 
 \frac{\nu^2 + \left( \frac{e^* V}{2 \pi \bT}  \right)^2}{2 \nu+1}
  \left\{  -2 \pi  \text{Im}~\psi \left( \nu + 1 + i \frac{e^* V}{2 \pi \bT} \right)
\tanh \left( \frac{e^* V}{2 \bT}\right) 
   \right. \nonumber \\
 & \qquad \qquad \qquad \left. + \frac{\pi^2}{2} +2 \left[ \text{Im}~\psi \left( \nu + 1 + i \frac{e^* V}{2\pi \bT} \right) \right]^2   -  \text{Re}~\psi' \left( \nu + 1 + i \frac{e^* V}{2 \pi \bT} \right) 
 \right\} 
\end{align}
which lead back to the expressions of Eqs.~\eqref{eq:fullcalS} and \eqref{eq:calC2nu} in the limit of vanishingly small voltage bias.

In the special case $\nu=1$, this can be worked out explicitly as
\begin{align}
 \cC_1^{(2)} (V) &=  \pi \frac{\pi  \frac{e^* V}{2 \pi \bT} \left[ 1 + \left( \frac{e^* V}{2 \pi \bT} \right)^2 \right]- \left[ 1 + 3\left( \frac{e^* V}{2 \pi \bT} \right)^2 \right] \tanh \left(  \frac{e^* V}{2  \bT} \right)}{3  \frac{e^* V}{2 \pi \bT} \left[ \sinh \left(  \frac{e^* V}{2  \bT} \right) \right]^2}
\end{align}
where we used that $\text{Im}~\psi \left( 2 + i z \right) = \frac{\pi}{2 \tanh (\pi z)} -\frac{1}{2z} - \frac{z}{1+z^2}$. 


\section{Crossed correlations and backscattered noise}

In the setup considered here, and represented in Fig.~1 of the main text, the current operators $I_\mu$ at the position of the contacts $x_\mu$ ($\mu = 3,4$) are given by
\begin{align}
I_3 (x_3, t) &= \frac{e \sqrt{\nu}}{2 \pi} v_F \partial_x \phi_R (x_3,t) + \frac{\nu e^2}{2 \pi} V \\
I_4 (x_4, t) &= - \frac{e \sqrt{\nu}}{2 \pi} v_F \partial_x \phi_L (x_4,t) .
\end{align}
We now focus on the weak backscattering regime, where the tunneling Hamiltonian is given by Eq.~\eqref{eq:HWB}. Expanding to second order in the tunneling parameter $\Gamma_0$, following a similar derivation to the one presented in the previous sections, one has for the average currents at the contacts
\begin{align}
\langle I_3 (x_3, t) \rangle &= 2i \frac{\nu e \Gamma_0^2}{(2 \pi a )^2} \int d\tau \sin \left( \nu e V \tau \right) \exp \left[ \nu \cG_R (-\tau) + \nu \cG_L (-\tau) \right] + \frac{\nu e^2}{2 \pi} V = - \cI_B  + \frac{\nu e^2}{2 \pi} V \\
\langle I_4 (x_4, t) \rangle &= - 2i \frac{\nu e \Gamma_0^2}{(2 \pi a )^2} \int d\tau \sin \left( \nu e V \tau \right) \exp \left[ \nu \cG_R (-\tau) + \nu \cG_L (-\tau) \right] = \cI_B ,
\end{align}
where we used the expression for the backscattered current obtained in Eq.~\eqref{eq:avgIB}.

The current crossed correlations measured at the contacts can similarly be defined as
\begin{align}
S_{34} (t-t') = \left\langle I_3(x_3, t) I_4 (x_4, t') \right\rangle - \left\langle I_3(x_3, t) \right\rangle \left\langle I_4 (x_4, t') \right\rangle
\end{align}
Substituting the expression for the current operators and performing an expansion to second order in the tunneling Hamiltonian, one obtains after some algebra
\begin{align}
S_{34} (t-t') &= - \left( \frac{e \nu v_F \Gamma_0}{4 \pi^2 a}  \right)^2 \sum_{\eta_1, \eta_2} \int d\tau \cos \left( \nu e V \tau \right)  \exp \left[ \cG_R (\sigma_{\eta_1 \eta_2} (\tau)) + \cG_L (\sigma_{\eta_1 \eta_2} (\tau)) \right] \nonumber \\
& \qquad \times \int d \bar{t} \frac{\pi T_R}{v_F} \left[ \coth \left( \pi T_R \left( \bar{t} - \frac{\tau}{2} - i \eta_1 \tau_0 \right) \right)  - \coth \left( \pi T_R \left( \bar{t} + \frac{\tau}{2} - i \eta_2 \tau_0 \right) \right) \right] \nonumber \\
& \qquad \qquad \times \frac{\pi T_L}{v_F} \left[ \coth \left( \pi T_L \left( \bar{t} - \frac{\tau}{2} -t+t' - i \eta_1 \tau_0 \right) \right)  - \coth \left( \pi T_L \left( \bar{t} + \frac{\tau}{2} -t+t'- i \eta_2 \tau_0 \right) \right) \right] .
\end{align}

Following the same prescription as the one used for the backscattered current, one can define the zero-frequency crossed correlations of the current as
\begin{align}
\cS_{34} &= 2 \int d\tau S_{34} (\tau) \nonumber \\
&= \left( \frac{e \nu \Gamma_0}{\pi a} \right)^2 \left\{ -  \int d\tau \cos \left( \nu e V \tau \right)  \exp \left[ \nu \cG_R (\tau) + \nu \cG_L (\tau) \right]   \right. \nonumber \\
& \qquad\qquad \qquad \left. + i \int d\tau \cos \left( \nu e V \tau \right) \left( T_R + T_L \right) \tau \exp \left[ \nu \cG_R (\tau) + \nu \cG_L (\tau) \right] \right\}
\end{align}
Using the expressions for the backscattered current and noise, Eqs.~\eqref{eq:avgIB} and \eqref{eq:zerofreqS}, one can readily write
\begin{align}
\cS_{34} &= 2 \left( T_R + T_L \right) \frac{\partial \cI_B}{\partial V} - \cS_B 
\end{align}
therefore providing a connection between tunneling conductance, crossed correlations and backscattered noise.


\section{Weakly coupled non-chiral Luttinger liquids}

In this section, we consider the slightly different case of two weakly coupled non-chiral Luttinger liquids. The full Hamiltonian of such a system is given by $H = H_1 + H_2 + H_T$ with
\begin{align}
H_\mu &= \frac{v}{2} \int dx \left[ K \left( \partial_x \phi_\mu \right)^2 +  \frac{1}{K} \left( \partial_x \theta_\mu \right)^2  \right] \\
H_T &= \Gamma \sum_{r, r'} \Psi_{r, 1}^\dagger (0) \Psi_{r', 2} (0)+ \text{H.c.}
\end{align}
where $v$ and $K$ are the standard Luttinger parameters, and $\Gamma$ is the tunneling constant. Note that we restrict ourselves to repulsive interactions, so that $K < 1$.

Following conventional notations, the bosonic modes $\phi_\mu$ and $\theta_\mu$ are related to the physical fermionic fields $\Psi_{r,\mu}$ via the bosonization identity
\begin{align}
\Psi_{r,\mu} (x) = \frac{U_{r,\mu}}{\sqrt{2 \pi a}} e^{i r k_F x} e^{i \sqrt{\pi} \left[ \phi(x) + r \theta (x) \right]}
\end{align}
where $U_{r,\mu}$ is a Klein factor, $a$ is a short distance cutoff, and $r = \pm 1$ corresponds to right/left movers.

The backscattered current operator is readily obtained from the tunneling Hamiltonian and reads
\begin{align}
\cI_B &= i e \Gamma \sum_{r, r'} \Psi_{r, 1}^\dagger (0) \Psi_{r', 2} (0)+ \text{H.c.}
\end{align}

The backscattered noise is similarly defined as
\begin{equation}
S_B (t,t') = \left\langle I_{B} (t) I_{B} (t') \right\rangle - \left\langle I_{B} (t) \right\rangle \left\langle I_{B} (t') \right\rangle
\end{equation}
and can be expanded to second order in the tunneling parameter $\Gamma$ yielding
\begin{align}
S_B (t,t') &= \left\langle T_K \Delta I_{B} (t^{(+)}) \Delta I_{B} (t'^{(-)}) \exp\left(  - i \sum_{\eta = \pm} \eta \int_{-\infty}^{+\infty} dt'' H_T (t''^{(\eta)}) \right)  \right\rangle \nonumber \\
&=  2 \left( \frac{e \Gamma}{2 \pi a} \right)^2 \sum_{r, r'}  \exp \left\{ \pi \left[ \cG_{\phi\phi,1}^{+-} \left( t,t' \right) + \cG_{\theta\theta,1}^{+-} \left( t,t' \right) + r \cG_{\phi\theta,1}^{+-} \left( t,t' \right) + r \cG_{\theta\phi,1}^{+-} \left( t,t' \right)  \right] \right\} \nonumber \\
& \qquad \qquad \qquad \qquad \times \exp \left\{ \pi \left[ \cG_{\phi\phi,2}^{+-} \left( t,t' \right) + \cG_{\theta\theta,2}^{+-} \left( t,t' \right) + r' \cG_{\phi\theta,2}^{+-} \left( t,t' \right) + r' \cG_{\theta\phi,2}^{+-} \left( t,t' \right)  \right] \right\}
\end{align}
where, for simplicity, we focused on the case of a temperature difference, setting the voltage bias $V=0$.

The Keldysh Green's functions are then readily expressed in terms of the conventional bosonic propagators $\cG_{\alpha\beta,\mu}^{+-} (t,t') = \cG_{\alpha\beta,\mu} (t'-t)$, whose expressions are obtained from standard Luttinger liquid derivations and are given by
\begin{align}
\cG_{\phi\phi,\mu} (\tau) &= - \frac{1}{2 \pi K} \log \left[ \frac{\sinh \left( \pi T_\mu (i\tau_0 - \tau) \right)}{\sinh \left( i \pi T_\mu \tau_0 \right)} \right] \\
\cG_{\theta\theta,\mu} (\tau) &= - \frac{K}{2 \pi} \log \left[ \frac{\sinh \left( \pi T_\mu (i\tau_0 - \tau) \right)}{\sinh \left( i \pi T_\mu \tau_0 \right)} \right] \\
\cG_{\phi\theta,\mu} (\tau) &= 0 \\
\cG_{\theta\phi,\mu} (\tau) &= 0
\end{align}
where $\tau_0 = a/v_F$ is a short time cutoff.

The zero-frequency backscattered noise then reads
\begin{align}
\cS_B &= 2 \int d\tau ~ S_B \left( \tau \right) \nonumber \\
& = 4 \left( \frac{e \Gamma}{\pi a} \right)^2  \int d\tau \exp \left\{ \pi \left[ \cG_{\phi\phi,1} (\tau) + \cG_{\theta\theta,1} (\tau) + \cG_{\phi\phi,2} (\tau) + \cG_{\theta\theta,2} (\tau) \right] \right\} \nonumber \\
& = 4 \left( \frac{e \Gamma}{\pi a} \right)^2  \int d\tau \left[ \frac{\sinh \left( i \pi T_1 \tau_0 \right)}{\sinh \left( \pi T_1 (i\tau_0 - \tau) \right)} \right]^{\frac{1}{2} \left( K + \frac{1}{K} \right)}  \left[ \frac{\sinh \left( i \pi T_2 \tau_0 \right)}{\sinh \left( \pi T_2 (i\tau_0 - \tau) \right)} \right]^{\frac{1}{2} \left( K + \frac{1}{K} \right)}
\end{align}
Interestingly, the resulting expression is similar (up to a trivial numerical prefactor) to the one obtained in the chiral case, Eq.~\eqref{eq:zerofreqS}, provided that one introduces an effective filling factor related to the Luttinger parameter $K$, namely
\begin{align} 
\nu_\text{eff} = \frac{1}{2} \left( K + \frac{1}{K} \right)
\end{align}
This allows us to use some of the results obtained in the chiral case in order to extend them to the non-chiral one. Most importantly, since $K <1$, this effective filling factor always satisfies $\nu_\text{eff} \geq 1$, thus leading to a \emph{positive} contribution to the delta-$T$ noise.

This derivation can easily be extended to include more degrees of freedom (spinfull Luttinger liquids, nanotubes) ultimately leading to similar expressions with an effective filling factor involving the multiple Luttinger liquid parameters, but still satisfying $\nu_\text{eff} \geq 1$.


\end{document}